# Dynamics of Self-retracting motion of graphite micro-flakes

Jiarui Yang[1,2], Yun Cao[3], Yao Cheng[1,3*]

[1]*Center for Nano and Micro Mechanics, Tsinghua University, Beijing 100084, China*

[2]*Department of Engineering Mechanics, Tsinghua University, Beijing 100084, China*

[3]*Department of Engineering Physics, Tsinghua University, Beijing 100084, China*

\* yao@tsinghua.edu.cn

**Abstract**

A sheared microscopic graphite mesa retracts spontaneously to minimize interfacial energy. Using an optical knife-edge technique, we report comparative measurements of the speeds of such self-retracting motion (SRM) in air and low vacuum (about 1 Pa). We found that the retracting speed $V_\mathrm{m}$ of SRM is almost the same in the two environments at both room temperature and high temperature (above 150 $^\mathrm{o}$C). We also extract the acceleration of the SRM, and found some typical behaviors. In the high speed range (above 5 m/s), the acceleration decreases linearly with the logarithm of the SRM speed. We attribute this phenomenon to the thermally activated sliding process in SRM. In the low speed region (between 0.1 m/s and 1 m/s), the acceleration increases linearly with the logarithm of the SRM speed.

**Introduction and Methods**

Q. Zheng, B. Jiang et al. observed self-retracting motion (SRM) of graphite micro-flakes [1] as shown in Fig. 1(b). The structure of the graphite mesa is shown in Fig. 1(a), with a silicon dioxide cover layer of thickness $h_1$, a graphite part of thickness $h_2$ and square shape of side length $L$. When a tungsten tip shears the mesa from the top, the graphite part is cleaved in the middle. After releasing the tip, the graphite flake above retracts spontaneously and realigns with the graphite mesa beneath to minimize the interfacial energy between them. Afterwards, J. Yang et al. measured the self-retracting motion of the graphite flakes by an optical knife edge method [2] as shown in Fig. 1(b), (c). A laser spot is focused on the edge of the retracting flake. The retraction will change the reflectivity of the area where the laser is on, therefore, the displacement of the retraction can be deduced by measuring the reflection and calibrate the intensity profile of the laser. Details of the knife-edge method is described in Ref [2].

Here, we measured the SRM with a higher signal to noise ratio (SNR) than the previous study and made comparative study of the speed of SRM in air and low vacuum environment (about 1 Pa). By analyzing the high SNR data, we also found that the acceleration during the SRM has a partially linear relation to the logarithm of the speed.

The method we used in this study is almost the same as in Ref [2]. We used Avalanche detector Hamamatsu C5331-03 with a lower sensitivity, and increase the input power of the laser to suppress the photon shot noise. For vacuum, we used a Varian TriScroll 300 dry Scroll Pump with an ultimate vacuum of 1.3 Pa.

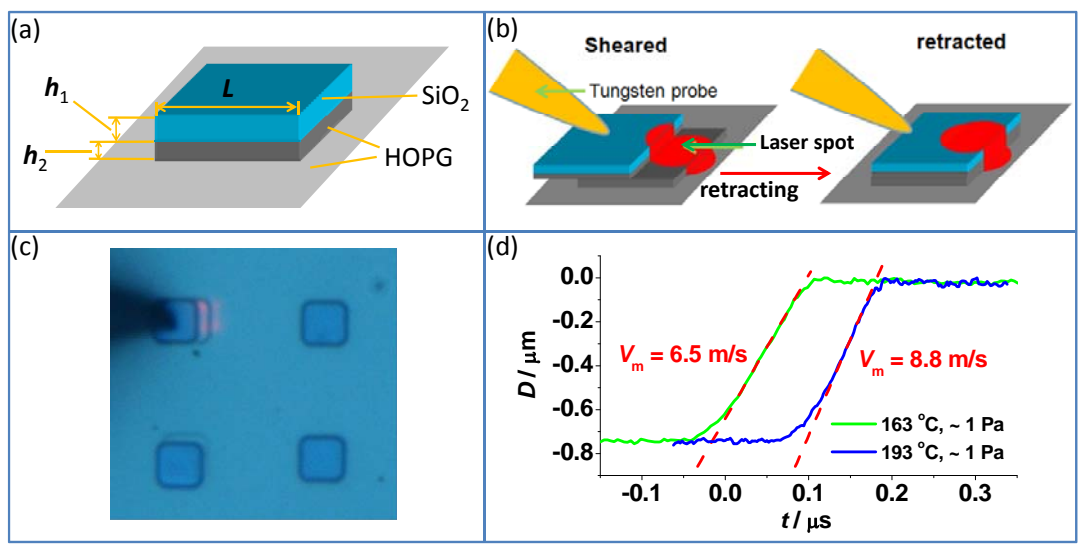

**FIG. 1** (a) Structure of the graphite mesa; (b),(c) shearing and retraction under optical microscope with a focused laser; (d) two typical retracting curves.

**Results:**

Two typical retracting motion curves measured on one mesa are shown in Fig. 1(d). The maximum speeds $V_m$ during the retraction are extracted to be 6.5 m/s and 8.8 m/s respectively. By using the $V_m$ as an indicator of friction during the retraction [2], we made comparison of $V_m$ in air and low vacuum as shown in Fig. 2. We can see the $V_m$ is almost not influenced by the vacuum. This is in contradiction with previous studies on the frictional properties of macroscopic graphite in vacuum [3-8]. We speculate this result comes from the fact that the graphite mesa we used in our experiment is almost single crystalline [9].

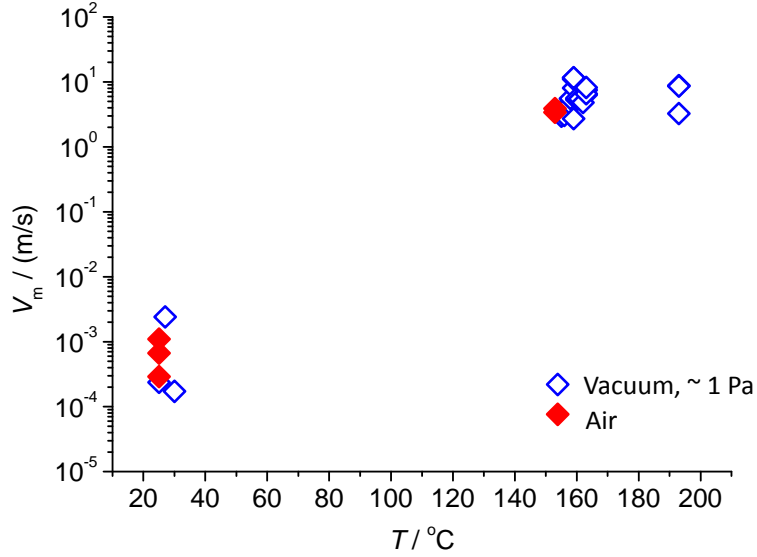

**FIG. 2** Comparison of the retracting speed $V_m$ in air and in low vacuum at different temperatures.

By extracting the velocity $V(t)$ and acceleration $A(t)$ from the displacement curves as in Fig. 1(d), we can make the plot $A - \log(V)$ as shown in Fig. 3. We used both normal average and cosine weighted average to calculate the $A(t)$ and $V(t)$ in these curves. The average time is 20 ns. We can see that during the retraction, while $V$ above 0.1 m/s, $A$ is linearly increasing with $\log(V)$ until $V$ reaches about 1 m/s. $A$ keeps almost constant when $V$ is between 1 m/s and 5 m/s, and starts to decrease linearly with $\log(V)$ after $V$ is above 5 m/s.

We already know that the friction at the sliding interface of the self-retracting graphite flakes is thermally activated [2]. And the nearly exponential dependence of $V_m$ on temperature $T$ is confirmed in our vacuum experiment as shown in Fig. 2. According to the thermally activated friction models [10-12], the friction $F_f$ depends on the logarithm of speed linearly. In our graphite retracting system, the retracting driving force is a constant $F_{ret}$ [1, 9], therefore the acceleration $A \propto F_{ret} - F_f$. is linear on $\log(V)$. As in the thermally activated friction model, $F_f$ increases linearly with $\log(V)$, therefore $A$ decreases linearly with $\log(V)$, which is consistent with the third section of the curve the in Fig. 3 while $V$ is above 5 m/s. In the first section, while $V$ is between 0.1 m/s and 1 m/s, the logarithmic dependence of $A$ on $V$ also indicates a thermal activation process, however has an opposite sign to the conventional friction model. Many similar results are given in reference [13]. The fine mechanism of the friction process in our graphite retracting system is not yet understood.

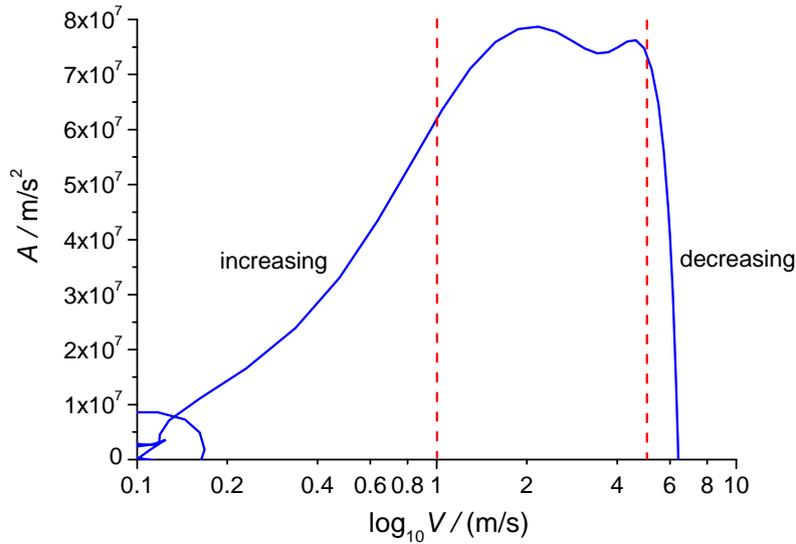

**FIG. 3** A typical plot of $A(t)$ to $\log(V(t))$ of SRM. The multiple values below 0.2 m/s was caused by the detecting noise.

**Conclusion**

We found that the retracting speed $V_m$ of SRM is almost the same in air and low vacuum at both room temperature and high temperature (above 150 $^o$C). The reason why vacuum has almost no influence to the frictional property of the graphite retracting system is due to the crystallinity of the graphite mesa. We also found that, in the high speed range during the SRM (above 5 m/s), the acceleration decreases linearly with the logarithm of the SRM speed. We attribute this phenomenon to the thermally activated sliding process in SRM. In the low speed region (between 0.1 m/s and 1 m/s), the acceleration increases linearly with the logarithm of the SRM speed , which has an opposite sign to the conventional models of thermally activated friction. The detailed mechanism of the friction process in SRM is not yet understood.

**Acknowledgement**

We thank professor Quanshui Zheng for his helpful discussions. We acknowledge the financial support from NSFC through Grants No. 10832005 and 973 Grants No. 2013CB934200.